\documentclass[letterpaper,twocolumn,10pt]{article}
\usepackage{usenix}
\usepackage{multicol}
\usepackage{hyperref}
\usepackage{tikz}
\usepackage{svg}
\usepackage{amsmath}
\usepackage{tabularx}
\usepackage{multirow}
\usepackage{comment}
\usepackage{filecontents}
\usepackage{algorithm} 
\usepackage{algpseudocode}  
\usepackage{algorithmicx}  
\usepackage{graphicx}  
\usepackage{array} 
\usepackage{caption}
\usepackage{afterpage}
\usepackage{threeparttable}
\usepackage{enumitem}
\usepackage{xspace}
\usepackage{todonotes}
\usepackage{wrapfig}
\usepackage{rotating}
\usepackage{lipsum}
\usepackage{booktabs}
\usepackage{todonotes}
\microtypecontext{spacing=nonfrench}

\begin{document}
%-------------------------------------------------------------------------------

%don't want date printed
\date{}

\definecolor{shangye}{RGB}{0, 128, 0}
\definecolor{dong}{RGB}{0,0, 180}
\definecolor{check}{RGB}{0,0,0}

%\title{\Large \bf An Empirical Guide to the Behavior and Use of CXL Memory}

\newcommand{\name}{Tuna\xspace}
\newcommand{\mycmdA}{Command A\xspace}

\title{\Large \bf \name: Tuning Fast Memory Size based on Modeling of Page Migration \\ for Tiered Memory}

%for single author (just remove % characters)
\author{%
\begin{minipage}[t]{.33\textwidth}
    \centering
    {\footnotesize \textbf{Shangye Chen}}\\
    \footnotesize University of California, Merced\\
    \texttt{\footnotesize schen251@ucmerced.edu}
\end{minipage}%
\begin{minipage}[t]{.33\textwidth}
    \centering
    {\footnotesize \textbf{Jin Huang}}\\
    \footnotesize University of California, Merced\\
    \texttt{\footnotesize jhuang326@ucmerced.edu}
\end{minipage}%
\begin{minipage}[t]{.33\textwidth}
    \centering
    {\footnotesize \textbf{Shuangyan Yang}}\\
    \footnotesize University of California, Merced\\
    \texttt{\footnotesize syang127@ucmerced.edu}
\end{minipage}
\\ % End of first row
\begin{minipage}[t]{.33\textwidth}
    \centering
    {\footnotesize \textbf{Jie Liu}}\\
    \footnotesize University of California, Merced\\
    \texttt{\footnotesize jliu279@ucmerced.edu}
\end{minipage}%
\begin{minipage}[t]{.33\textwidth}
    \centering
    {\footnotesize \textbf{Huaicheng Li}}\\
    \footnotesize Virginia Tech\\
    \texttt{\footnotesize huaicheng@cs.vt.edu}
\end{minipage}%
\begin{minipage}[t]{.33\textwidth}
    \centering
    {\footnotesize \textbf{Dimitrios Nikolopoulos}}\\
    \footnotesize Virginia Tech\\
    \texttt{\footnotesize dsn@vt.edu}
\end{minipage}
\\ % End of second row
\begin{minipage}[t]{.33\textwidth}
    \centering
    {\footnotesize \textbf{Junhee Ryu}}\\
    \footnotesize SK Hynix\\
    \texttt{\footnotesize junhee.ryu@sk.com}
\end{minipage}%
\begin{minipage}[t]{.33\textwidth}
    \centering
    {\footnotesize \textbf{Jinho Baek}}\\
    \footnotesize SK Hynix\\
    \texttt{\footnotesize jinho.baek@sk.com}
\end{minipage}%
\begin{minipage}[t]{.33\textwidth}
    \centering
    {\footnotesize \textbf{Kwangsik Shin}}\\
    \footnotesize SK Hynix\\
    \texttt{\footnotesize kwangsik.shin@sk.com}
\end{minipage}
\\ % End of third row
\begin{minipage}[t]{.33\textwidth}
    \centering
    {\footnotesize \textbf{Dong Li}}\\
    \footnotesize University of California, Merced\\
    \texttt{\footnotesize dli35@ucmerced.edu}
\end{minipage}
}

\maketitle

%-------------------------------------------------------------------------------
\begin{abstract}
%-------------------------------------------------------------------------------
Tiered memory, built upon a combination of fast memory and slow memory, provides a cost-effective solution to meet ever-increasing requirements from emerging applications for large memory capacity. Reducing the size of fast memory is valuable to improve memory utilization in production and reduce production costs because fast memory tends to be expensive. However, deciding the fast memory size is challenging because there is a complex interplay between application characterization and the overhead of page migration used to mitigate the impact of limited fast memory capacity. 
%%\dsn{This is a little vague, I would rephrase that there is a complex interplay between characterizing the sensitivity of applications against the available fast memory capacity and the overhead of page migration used to mitigate the impact of limited fast memory capacity (or a shorter version of this sentence).} 
In this paper, we introduce a system, \name, to decide fast memory size based on modeling of page migration. \name uses micro-benchmarking to model the impact of page migration on application performance using three metrics. \name decides the fast memory size based on offline modeling results and limited information on workload telemetry.  Evaluating with common big-memory applications and using 5\% as the performance loss target, we show that \name in combination with a page management system (TPP~\cite{10.1145/3575693.3578835}) \textcolor{check}{saves fast memory by 8.5\% on average (up to 16\%). This is in contrast to 5\% saving in fast memory reported by Microsoft Pond~\cite{10.1145/3575693.3578835} for the same workloads (BFS and SSSP) and the same performance loss target.} 

%Pond (a sate- focusing on untouched memory without page migration)  
\end{abstract}
\vspace{-10pt}
%%%\dsn{I would mention that Poind is a state-of-the-art system for managing tiered memory.}}.

%\subsection{LaTeX-ing Your TeX File}
%-----------------------------------

%People often use \texttt{pdflatex} these days for creating pdf-s from
%tex files via the shell. And \texttt{bibtex}, of course. Works for us.

%-------------------------------------------------------------------------------
%\section*{Acknowledgments}
%-------------------------------------------------------------------------------

%The USENIX latex style is old and very tired, which is why
%there's no \textbackslash{}acks command for you to use when
%acknowledging. Sorry.

%-------------------------------------------------------------------------------
%\section*{Availability}
%-------------------------------------------------------------------------------

%USENIX program committees give extra points to submissions that are backed by artifacts that are publicly available. If you made your code or data available, it's worth mentioning this fact in a dedicated section.

%-------------------------------------------------------------------------------

\section{Introduction}
\label{sec:intro}
\vspace{-5pt}
Many emerging applications call for memory with a larger capacity.  However, large DRAM capacity contributes significantly to server cost (e.g., 37\% of Meta server cost~\cite{10.1145/3582016.3582063}), because of the scaling limitation of DRAM technologies. With the rapid development of memory-intensive applications (e.g., AI training and inference~\cite{pmlr-v202-sheng23a,DBLP:journals/corr/abs-2104-07857,atc21:zerooffload} and HPC applications~\cite{ics21:warpx,ics21:athena}), existing memory systems face challenges in providing support for large DRAM capacity. 
\vspace{-1pt}
Memory tiering is a solution to address the above problem.  Based upon cache-coherence memory interconnect (e.g., Compute Express Link or CXL~\cite{cxl}) and non-DRAM memory technologies (e.g., non-volatile memory), memory tiering uses multiple types of memory with different capacities, latency, and monetary cost, providing a cost-effective solution to increase overall memory capacity. Tiered memory is often characterized by a fast memory tier with a small capacity and a slow memory tier with a large capacity. Tiered memory must rely on a page management system that decides page allocation at different tiers and migrates pages between tiers such that frequently accessed pages (i.e., hot pages) can usually be located in fast memory for high performance.

There is a strong need to reduce the usage of fast memory in production~\cite{Agarwal:2017:TAP:3037697.3037706,10.1145/3575693.3578835} in order to improve memory utilization in production and save production cost (because fast memory tends to be expensive). However, deciding the fast memory size is challenging because there is a complex interplay between the sensitivity of applications to fast memory capacity, and the overhead of page migration, which is a mechanism to alleviate the performance impact of limited fast memory capacity. Although most efforts for tiered memory focus on memory profiling and effective page migration, saving fast memory is often not an optimization target. Recent work on Pond~\cite{10.1145/3575693.3578835}, a system within Microsoft Azure, predicts untouched memory in the application and offloads it to slow memory within a performance-loss constraint, with the objective of saving fast memory. The prediction is based on machine learning models correlating workload telemetry, virtual machine metadata, and a minimum amount of untouched memory. This method assumes advanced knowledge of customer workloads. This knowledge may not be generally available in other public settings. Also, Pond does not have a page management system in place and does not support page migration between memory tiers. 

In this paper, we use page migration to further save fast memory capacity. We decide on the best fast memory size for a given application based on a general characterization and modeling of the impact of page migration on application performance. In particular, our motivation study reveals that using page migration, we can save fast memory size by 10.5\%, while without page migration we save at most \textcolor{check}{2.5\%}, under the same performance loss constraint (5\% as in Pond). Furthermore, we use three metrics to model the impact of page migration on application performance: number of page accesses in slow and fast memories, number of page migrations, and arithmetic intensity in the application. The three metrics capture (1) the competition between page migration and page accesses on memory bandwidth, (2) page migration overhead, and (3) how sensitive the application is to memory performance. We use a micro-benchmark to exercise metrics with various fast memory sizes offline. Those benchmarking results are collected into a performance database, providing guidance to determine fast memory size at runtime in production. This modeling method does not assume the availability of a priori knowledge of the applications and customer history; it is thus both feasible and general. 

We summarize the major contributions as follows.
\begin{itemize}[leftmargin=*,noitemsep,topsep=0pt]
\item We reveal the potential of saving fast memory with page migration in place, and introduce a new modeling method to decide fast memory size;

\item We design \name, a system that generates workloads for modeling of page migration and a runtime system to tune fast memory size;

\item Evaluating with common big memory applications, we show that \name in combination with TPP saves fast memory \textcolor{check}{by 8.5\% on average (up to 16\%). This is in contrast to 5\% saving in fast memory reported by Microsoft Pond~\cite{10.1145/3575693.3578835} for the same workloads (BFS and SSSP) and the same performance loss target.}  

\end{itemize}
\vspace{-15pt}
%which is 3.5\% more than Pond (a state-of-the-art solution focusing on untouched memory without page migration)

\section{Background}
\label{sec:bg}
\vspace{-5pt}
\textbf{Use scenario of tiered memory.} Our study is based on the following common use scenario: the whole memory footprint of the application is allocated in fast memory, but some memory pages can be offloaded to slow memory when they are not accessed or less frequently accessed. %guided by a memory profiling mechanism. 
By offloading memory pages, the system can save fast memory, but should not lose performance too much (compared with the performance of using fast memory only). The performance loss threshold $\tau$  is given by the user (e.g., $\tau = 5\%$ as in~\cite{10.1145/3575693.3578835}). This use scenario is aligned with the recent scenario in a data center~\cite{10.1145/3575693.3578835}.  

% \vspace{-3pt}
There is a page management system for tiered memory (e.g., TPP~\cite{10.1145/3582016.3582063}, HeMem~\cite{10.1145/3477132.3483550}, and MEMTIS~\cite{10.1145/3600006.3613167}). This system periodically migrates pages between fast memory and slow memory. This period is called the page migration interval. The page migration decision is based on page access frequency in one or multiple time intervals (called profiling intervals), such that frequently accessed pages are in fast memory and others are in slow memory. 
% \vspace{-3pt}
\textbf{Motivation.} To motivate our work, we run BFS in a two-tiered memory system based on Intel Optane (Section~\ref{sec:eval} has hardware details and workload details). The fast memory size is initially set to 12.4 GB, which is the peak memory consumption of BFS. We deploy TPP as a page management system in our evaluation. We reduce the fast memory size and study the performance of BFS. 

\begin{figure}    
\centering     
\includegraphics[width=0.45\textwidth]{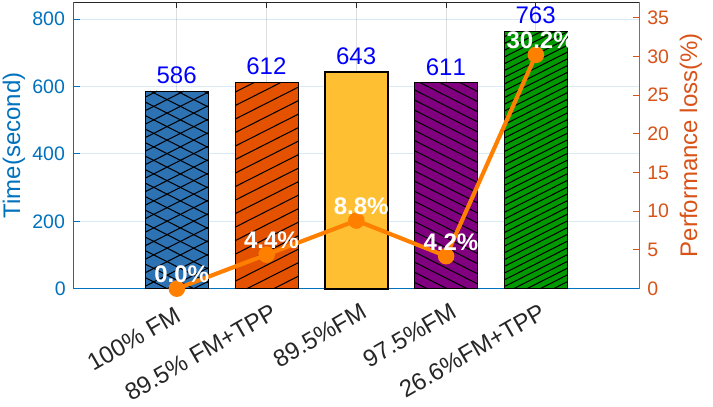}
\vspace{-10pt}
\caption{BFS performance with various fast memory sizes and a page management system (TPP) in place. Percentage numbers in the $x$ axis are the percentage of the original fast memory size. ``FM'' stands for fast memory.}
\label{fig:motivation}
\end{figure}

Figure~\ref{fig:motivation} shows the results. The performance is best when allocating all pages in fast memory, a setup labeled ``fast memory only''. When the fast memory size is reduced to  89.5\%, without the page management system, BFS has a 8.8\% performance loss, compared to the performance of fast memory only. In this case, BFS uses a NUMA first-touch policy: the pages are allocated to fast memory first, and when the fast memory is full, new pages are spilled to slow memory. Since there is no page migration, the hot pages may be allocated to slow memory. With TPP and the same fast memory size, BFS avoids such a large performance loss: the performance loss is 4.4\%. Meanwhile, the system saves 10.5\% of fast memory. However, when the fast memory becomes even smaller (i.e., 26.6\% of the fast memory size), BFS has a large performance loss (30.2\%), even with the page management system in place. With such a small fast memory size, the frequent page movement and inability to host hot pages is detrimental to performance. In fact, compared with using 89.5\% of fast memory, using 26.6\% of fast memory increases the page migration failures by 21\% and the number of page migrations by 40\%. The page migration failure happens when TPP promotes hot pages from slow memory to fast memory while fast memory does not have enough space. The page reclaim in the fast memory happens in the background when there is no enough fast memory space, but cannot capture up with the rate of page promotion, leading to page migration failures. 
%\dsn{Define what is a page migration failure.

The above results demonstrate the importance of finding an appropriate fast memory size without causing large performance loss.

% \vspace{25pt}
\vspace{-5pt}
%%%\textcolor{red}{Discuss the impact of fast memory size on application performance.}

%%%%\textcolor{red}{Observation 1: (1) reducing the fast memory size does not lead to performance loss, when the page migration is in place (but when the page migration is not there, there will be performance loss).  (2) reducing the fast memory size does not lead to perf loss when the page migration is not there. (3) page migration can lead to performance loss.}

%\input text/overview
\vspace{-5pt}
\section{Modeling Page Migration}
\label{sec:sys_abs}
\vspace{-5pt}
%\textcolor{shangye}{ Figure \ref{fig:overview} shows \name, which is the whole structure of our design and model. There are two parts in total, one is offline and another is online. Tuna helps the model to communicate between online and offline. The details of the design is shown as follows. }

\begin{figure}    
\centering     
\includegraphics[width=0.45\textwidth]{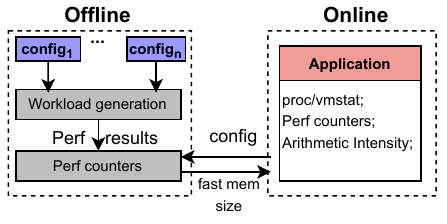}
\vspace{-12pt}
\caption{Overview of \name}
\label{fig:overview}
\end{figure}

Figure \ref{fig:overview} overviews \name. We depcit how to model the impact of page migration on application performance in this section. We characterize the interaction between page migration and application performance as follows.
%which is the whole structure of our design and model. There are two parts in total, one is offline and another is online. Tuna helps the model to communicate between online and offline. The details of the design is shown as follows. 

% \vspace{-10pt}

%Our modeling is based on the following use scenario: the whole memory footprint of the application is allocated in fast memory, but some memory pages can be offloaded to slow memory when they are less frequently accessed, guided by a memory profiling mechanism. By offloading memory pages, the system can save fast memory, but should not lose performance too much (compared with the case of fast memory only). The performance loss threshold $\tau$  is given by the user based on QoS (e.g., $\tau = 5\%$ as in~\cite{10.1145/3575693.3578835}). This use scenario is aligned with the real scenario in a data center~\cite{10.1145/3575693.3578835}.  

%We assume that there is a page management system in the above use scenario (e.g., TPP, Thermostat, and Memtis). This system periodically migrates pages between fast memory and slow memory. This period is called the page migration interval. The page migration decision is based on page access frequency in one or multiple time intervals (called profiling intervals).  

\begin{itemize}[leftmargin=*,noitemsep,topsep=0pt]
    \item The page migration competes with the application for memory bandwidth. If the application performance is bounded by memory bandwidth, a high page migration traffic can lead to large performance loss in the application. 

    \item The application performance can be insensitive to long memory access latency or low memory bandwidth. The performance insensitivity can come from high computation intensity and from using processor architectures with high instruction-level or thread-level parallelism. 
\end{itemize}

%To study the impact of page migration on the application performance, we emulate the above interactions using a micro-benchmark. 

When the availability of fast memory capacity changes, page migration frequency also changes and affects application performance. We emulate the above interaction using a micro-benchmark to decide the minimum fast memory size without violating the constraint on performance loss. 
%\vspace{-3pt}
\begin{comment}
\subsection{Characterization of Page Migration} 
From the perspective of page migration, the micro-benchmark should share three characterizations with the application. Within a profiling time interval, %%(denoted by $prof\_int$), 
compared with the application, (1)  the micro-benchmark causes the same number of page migrations between fast memory and slow memory, (2) has the same arithmetic intensity, characterized as attainable floating point operations and integer operations per second (i.e., FLOPS and IOPS), and (3) has the same number of page accesses.

The characterization (1), denoted by $pm$ (including $pm\_de$ and $pm\_pr$ for page demotion and promotion respectively), is necessary to reflect the page migration overhead. The characterizations (2) (denoted by $AI$) and (3) (denoted by $pacc\_f$ and $pacc\_s$ for fast and slow memories respectively) are necessary to model the application sensitivity to memory performance and memory bandwidth consumption from the application respectively.  
\end{comment}

\vspace{-5pt}
\subsection{Characterization of Page Migration} 
\vspace{-5pt}
From the perspective of page migration, the micro-benchmark should share three characteristics with the application. Within a profiling time interval, %%(denoted by $prof\_int$), 
compared with the application, the micro-benchmark should have (1) the same number of page migrations between fast memory and slow memory, (2) the same arithmetic intensity, characterized as attainable floating point operations and integer operations per second (i.e., FLOPS and IOPS), and (3) the same number of page accesses.
%\dsn{Should the microbenchmark be identical to the application in these characteristics or approximate them as best as possible?}

Performance characteristic (1), denoted by $pm$ (including $pm\_de$ and $pm\_pr$ for page demotion and promotion, respectively), is necessary to reflect the page migration overhead. Characteristic (2) denoted by $AI$ and characteristic (3) denoted by $pacc\_f$ and $pacc\_s$ for fast and slow memories respectively are necessary to model the application sensitivity to memory performance and memory bandwidth consumption, respectively.

\vspace{-5pt}
\subsection{Micro-benchmark Design}  
\vspace{-5pt}
%\dsn{Have you defined RSS before? It might be self-explanatory, but best to define everything.}
Taking $pacc$, $pm$, $AI$, %%%$prof\_int$ 
and the application's resident set size (RSS) as input, the micro-benchmark generates a workload. In essence, the workload performs strided memory accesses to two arrays (one in fast memory and the other in slow memory) whose physical memory consumption equals RSS. By mapping array elements to pages, manipulating the stride size to control the number of memory accesses per page, and limiting computation per memory access, we reach the goals of $pacc$, $pm$, and $AI$.

The micro-benchmark uses the strided memory accesses, because this access pattern allows us to easily bypass the cache hierarchy and control the number of memory accesses at the page level. Note that the micro-benchmark is not designed to follow the memory access patterns of the application exactly. Rather, the micro-benchmark aims to emulate page accesses and migration. From the perspective of page migration on tiered memory, $pacc$, $pm$, and $AI$ determine the performance on tiered memory, and we focus on their impacts. There is no need to model how memory accesses happen (e.g., either using random, stencil, or strided memory access patterns). 
%%%\dsn{This is a little confusing. I thought you control the stride to emulate certain behavior, but this sentence implies that the application can have a non-strided access pattern that behaves like a strided access pattern. Is this right?}

%Taking $pacc$, $AI$, and RSS as input, the micro-benchmark generates a workload. In essence, the workload performs strided memory accesses to two arrays (one in fast memory and the other in slow memory) whose physical memory consumption are equal to RSS. By mapping array elements to pages, manipulating the stride size to control the number of memory accesses per page, and limiting computation per memory access, we reach the goals of $pacc$ and $AI$.

%The micro-benchmark uses the strided memory accesses, because this access pattern allows us to easily bypass the cache hierarchy and control the number of memory accesses at the page level. Note that the micro-benchmark is not designed to exactly follow the memory access patterns of the application. Rather, the micro-benchmark aims to emulate page accesses and migrations. From the perspective of page migration on tiered memory, $pacc$, $AI$ and page migration determine the performance on tiered memory, and we focus on their impacts and there is no need to model how memory accesses happen (e.g., either using random, stencil, or strided memory access patterns). 

\textbf{Meeting the requirement on \textit{pacc}.} 
The micro-benchmark allocates $NP\_fast$ number of pages and $NP\_slow$ number of pages in fast and slow memories, respectively ($NP\_slow = 0$ for fast memory only). These pages are accessed once to be physically allocated, in the initialization phase of the micro-benchmark. After that, the micro-benchmark starts the workload. %executed with the page management system in place. 

Given the page promotion threshold in the page management system (i.e.,  $hot\_thr$, the number of memory accesses in a page that can trigger page promotion), $NP\_fast$ and $NP\_slow$ are defined in Equations~\ref{eq:np_fast} and~\ref{eq:np_slow} respectively. These definitions do not include page accesses due to page migration. The pages counted in the two equations are accessed $hot\_thr - 1$ times each. Page accesses due to page migration are not considered in the two equations but are considered to meet the requirement on $pm$. Each promoted page is accessed $hot\_thr$ times, and each demoted page is accessed once.
\vspace{-10pt}

\begin{align} 
pacc\_fast &= pacc\_fast - pm\_de \times 1 \label{eq:pacc_fast}\\
pacc\_slow &= pacc\_slow - pm\_pr \times hot\_thr \label{eq:pacc_slow}\\
NP\_fast &= pacc\_fast \; / \; hot\_thr \label{eq:np_fast}\\
NP\_slow &= pacc\_slow \; / \; (hot\_thr-1)  \label{eq:np_slow}
\end{align}
\vspace{-1pt}
$hot\_thr$ is invariant in some page management systems (e.g., TPP\cite{10.1145/3582016.3582063}, AutoNUMA\cite{autonuma}, AutoTiering\cite{atc21:autotiering}, and M3~\cite{eurosys24:m3}). But other page management system (e.g., MEMTIS)  dynamically fine-tune $hot\_thr$ according to the availability of fast memory. For such a dynamic  $hot\_thr$, its value is given as an input when the runtime queries the performance database (see Section~\ref{sec:perf_db}).

\textbf{Meeting the requirement on \textit{pm}.} The micro-benchmark allocates $pm\_pr$ and $pm\_de$ number of pages in slow memory and fast  memory, respectively. Each of those $pm\_pr$ pages in slow memory is accessed $hot\_thr$ times, and then promoted to fast memory. Each of those $pm\_de$ pages in fast memory is accessed once and then demoted to slow memory. Page promotion and demotion happen randomly during the workload execution. 

%To reach of the goal of $pm\_de$, we must consider how page demotion happens. In most of the page management systems, the page demotion (or page reclaim) happens when the free memory space is below a threshold ($free\_page\_thr$). The page demotion is decoupled from the page promotion. To trigger the desired page demotion, the fast memory size is initially set as ($NP\_fast$ + $NP\_slow$ + $pm\_de$ + $pm\_pr$ + $free\_page\_thr$), but is reduced by $pm\_pr$ after all pages are accessed once. The control of fast memory size is implemented by co-running another micro-benchmark which does nothing but allocates physical pages in fast memory to control its available size. 

\textbf{Meeting the requirement on \textit{AI}.} A page access is implemented by loading an array element from main memory. Once the element is loaded, the micro-benchmark performs random floating-point multiplications and integer additions to meet the requirement on AI.

%%%%%\textbf{Meeting the requirement on \textit{prof\_int}.} The micro-benchmark starts a timer to measure the time for page migration, performing arithmetic operations, and memory accesses. The time should be no larger than \textit{prof\_int}, because the micro-benchmark, although performing the same page accesses, does not have overhead (e.g., thread synchronization or control flow) as the application. 

\textbf{Limitation.} 
In the micro-benchmark, memory accesses are evenly spread across pages in slow and fast memories, thus maximizing memory level parallelism. In the real application, memory accesses may not be evenly distributed and can be serialized to the same page, which leads to worse performance. The micro-benchmark does not model such performance, and hence models the best memory performance. 
\vspace{-5pt}
%%%\textbf{Limitation.} In (1), we do not distinguish fast-to-slow memory migration and slow-to-fast memory migration, based on the assumption that the migrations in two direction share the same memory bandwidth. But some memory (such as Intel Optane as slow memory) has performance asymmetry in load and store accesses. As a result, the migrations in two directions have different bandwidths. To handle this migration performance difference, we can extend our micro-benchmark by differentiating page promotion and demotion. \textcolor{red}{(to discuss more.)}
\vspace{-5pt}
\subsection{Building and Using Performance Database} 
\label{sec:perf_db}
\vspace{-5pt}
The performance database collects the execution records of the micro-benchmark. Each execution record is a set of execution times of the micro-benchmark using various fast memory sizes under a specific configuration. The configuration is an eight-element vector, shown as follows. The execution record is indexed using the configuration. 
%\vspace{5pt}

\footnotesize 
 [$pacc\_f$, $pacc\_s$, $pm\_de$, $pm\_pr$, $AI$, RSS, $hot\_thr$, num\_threads]
\normalsize

%\vspace{5pt}
The configuration includes the number of threads (num\_threads) to run the micro-benchmark. When multi-threading is used, each thread respects $AI$, and evenly shares $pm$ and $pacc$. The performance database is built offline by exercising various fast memory sizes for each configuration. 

Once the performance database is built, it can be used to tune the fast memory size at runtime. The runtime periodically measures $pacc$, $pm$, $AI$, RSS, and num\_threads to compose a configuration vector to query the performance database. The execution record whose index is closest to the input configuration vector is returned from the performance database. 
%\dsn{What metric do you use to compute the ``closeness'' of the application to the configuration vector?} 
The runtime uses the execution record to choose a minimum fast-memory size whose performance is within a performance loss target specified by the user. If there is no such a size, the runtime keeps the current fast memory size. 

Note that when using the performance database, the performance loss is calculated using the micro-benchmark performance (not the application performance) measured with \textit{fast memory only} as the baseline. This method improves the modeling accuracy because the baseline performance and the performance with smaller fast memory are collected using the same application (i.e., the micro-benchmark). In contrast, using the application to collect the baseline and using the micro-benchmark to collect the performance with smaller fast memory size, we cannot effectively save fast memory size, because the micro-benchmark only focuses on modeling the impact of page migration. 

\vspace{-10pt}

\vspace{-1pt}
\section{Tuning Fast Memory Size}
\label{sec:tuning}
\vspace{-5pt}

\textbf{Decreasing fast memory size} relies on page reclamation mechanisms in the operating system (OS) and page management system to demote pages from fast memory, such that the application can only utilize a limited fast-memory size. Once memory consumption becomes larger than the size, page reclamation kicks in. 

In Linux, there are three thresholds to determine when and how page reclamation should happen: min watermark, low watermark, and high watermark. Direct memory reclamation happens when the free memory size is below the min watermark. This reclamation mechanism blocks the application process, which can cause performance loss. When the free memory size is below the low watermark, a kernel thread \texttt{kswapd} is activated to start memory reclamation until the high watermark is reached. \texttt{kswapd} operates independently and asynchronously in the background, and does not block the application. 

We want to use \texttt{kswapd} instead of direct page reclaim to minimize the impact of decreasing fast memory size on the application. Assuming we want to reduce the fast memory size to $new\_fm$, the low watermark should be set to $new\_fm$. In Linux, the low watermark and min watermark are tightly coupled ($min\_watermark \approx 0.8 \times low\_watermark$~\cite{Rik,Torvalds_1991}),  such that the min watermark is always smaller than the low watermark, and setting the low watermark is achieved by setting the min watermark. We set the min watermark to $0.8\times new\_fm$. Linux uses a complex ``if condition'' to decide if page reclamation should happen: both the watermarks and the free memory size are used to make the decision. To enable watermark-based control of the fast memory size, we simplify the condition such that the watermarks are the only criteria to trigger page reclamation.  

Once page reclamation happens, it will continue until the free memory size reaches the high watermark. To ensure that the kernel does not claim too many pages and the fast memory size is smaller than $new\_fm$, we set the high watermark to $new\_fm$. 

%\textcolor{red}{After direct memory reclamation, if the remaining free memory in the system is still insufficient for memory allocation, then the Out of Memory (OOM) will be further triggered. The OOM Killer mechanism will terminate a process with high physical memory usage based on the algorithm to free up memory resources. If the physical memory remains insufficient even after killing the process, the OOM Killer will continue to terminate processes with high physical memory usage until enough memory is released.}

\textbf{Increasing fast memory size} is implemented by controlling the watermarks in the same way as decreasing fast memory size. Once the watermarks are changed, the page promotion failures due to the limited fast memory size are reduced, and the page management system can migrate more pages to the fast memory, effectively utilizing increased fast memory space. 
\vspace{-12pt}

\vspace{-5pt}
\section{Implementation}
\label{sec:impl}
%%\textcolor{red}{(discuss how to measure pacc, pm and AI using performance counters. discuss how to change the page relcaim threshold. Discuss how many records are there in the performance database. Discuss how long to build it)}
\vspace{-5pt}
\name has two components: an offline component for building the performance database, and an online component for fine-tuning the fast memory size. The offline component includes (1) a micro-benchmark template with $pacc$,$pm$, $AI$, and RSS as variables. Once instantiated, the micro-benchmark becomes a workload that emulates the application on tiered memory. The offline component also includes (2) a performance database based on Faiss~\cite{faiss} where configuration vectors along with execution times of the micro-benchmarks with various fast memory sizes are stored.  The configuration vectors are structured into a hierarchical graph in Faiss for quick search. We collected 100K execution records (100K configuration vectors) and each record has 100 workload-execution times with various fast memory size. Indexing those execution records takes us less than 20 minutes. Since each configuration vector is low-dimensional, a query of the database takes 500 $\mu s$. 

The online component includes a runtime library and minor changes to the kernel to control page reclaim. The minor kernel change is used to simplify the condition to trigger page reclaim (see Section~\ref{sec:tuning}). The runtime library is linked with the application and profiles the application. The runtime library uses performance counters to measure AI and the number of memory accesses, uses the proc file system to count the number of page migrations (particularly, \texttt{/proc/vmstat}), and builds configuration vectors. The runtime library then queries the performance database and changes the fast memory size by changing the page reclaim watermarks (\texttt{/proc/sys/vm/min\_free\_kbytes} and \texttt{/proc/sys/vm/high\_free\_kbytes}). %%%\textcolor{red}{(double check the above sentence.)}
\vspace{-12pt}

%\textcolor{shangye}{We implemented our design in Linux Kernel v5.13.0-rc6+. We employ Linux's Perf tool to measure arithmetic intensity (AI). The measurement of $pacc$ is done through the performance counter by running the events in PEBS (Precise Event-Based Sampling). The $pm$ value is obtained by analyzing the system file $/proc/vmstat$, the relevant field for $pm$ is $'numa\_pages\_migrated'$, which represents the number of pages that have been migrated. The page reclaim threshold is conducted in the Linux system file of $/proc/sys/vm/min\_free\_kbytes$. There are more than 1610 records in our current performance database. The total time for building the performance database is around 80 hours.}

\section{Evaluation}
\label{sec:eval}
\vspace{-5pt}

%%\textcolor{red}{benchmarks: BFS, SSSP, XSBenchmark, pagerank, btree.}

\begin{comment}
\textcolor{red}{(1) Performance prediction accuracy based on the micro-benchmarks for five benchmarks;
(2) Given a user-specified performance loss target, 5\%, fast memory size variance and performance variance ---> save fast memory size for one or two benchmarks. Show the results with a runtime figure. (3) page migration at runtime. Compare TPP with our solution. Show the results with either a table or a runtime figure. (4) sensitivity study. (a) change the user-specified performance loss target, 5\%, 10\%, 15\% ---> no matter how we change the threshold, the application performance is always within the constraint. Showing one benchmark is enough. (b) xxxxx.}
\end{comment}
\begin{comment}
\begin{figure}    
\centering     
\includegraphics[width=0.5\textwidth]{figures/evaulation2xsbench.pdf}
%%\caption{Fast memory variance and performance accuracy}
\label{fig3:fastmemoryvariance}
\end{figure}

\begin{figure}    
\centering     
\includegraphics[width=0.5\textwidth]{figures/evaluation2BFS.pdf}
%%\caption{Fast memory variance and performance accuracy}
\label{fig4:fastmemoryvariance}
\end{figure}
\end{comment}
\vspace{-5pt}
\textbf{Evaluation platform}
We use a two-socket machine. Each socket has an Intel Xeon Gold 6252 CPU (24 cores), 756GB Intel Optane DC PM, and 96GB DRAM. Each CPU has local DRAM as fast memory and Optane as slow memory. Our evaluation uses one socket. 

%%\textcolor{shangye}{Our evaluation platform is based on a two-socket machine. Each socket has an Intel Xeon Gold 6252 CPU (24 cores), 756GB Intel Optane DC PM, and 96GB DRAM. There are in total four memory components and two PM components appear as CPU-less memory nodes in Linux. Each CPU has its local DRAM and slow memory which is a PM component. To eliminate interference from other factors, we conducted the evaluations using only one single socket.  }

\begin{table}
   \small
    \caption{Workload description.}
    \vspace{-10pt}
   \begin{tabularx}{\linewidth}{p{1.5cm} p{1cm} >{\small}X}
  
     %% \cline{1-3}
     \hline\hline
      \textbf{Workload} & \textbf{RRS} & \textbf{Description}                            \\ %%\cline{1-3} 
      \hline\hline
      PageRank           & 15.8 G      & Compute PageRank score \cite{beamer2015gap}                        \\ \cline{1-3}
      XSBench            & 16.4 G     & Monte Carlo neutron transport algorithm computation \cite{tramm2014xsbench}\\ \cline{1-3}
      BFS                & 12.4 G      & Compute with Breadth-First Search algorithm \cite{beamer2015gap}     \\ \cline{1-3}
      SSSP               & 23.5 G     & Compute with Single-Source Shortest Path algorithm \cite{beamer2015gap}       \\ \cline{1-3}
      Btree              & 10.8 G     & Retrieve data by in-memory index \cite{btree}      \\ \cline{1-3}
   \end{tabularx}
    
    \label{tab:benchmark}
\end{table}
\vspace{-1pt}
\textbf{Workloads.} We use representative, memory-consuming workloads for evaluation, including graph-related benchmarks (BFS, SSSP, and PageRank), an HPC benchmark (XSBench), an in-memory index (Btree). These workloads are commonly used for evaluating tiered memory systems\cite{10.1145/3582016.3582063,10.1145/3600006.3613167,kumar2021radiant,panwar2021fast,alverti2020enhancing,achermann2020mitosis,ram2021trident,panwar2019hawkeye}. Table \ref{tab:benchmark} summarizes these workloads. \textcolor{check}{We initially set the fast memory size to the peak memory consumption of each workload using GRUB memory map. The fast memory saving reported in this section is in terms of the peak memory consumption of the workload.} 
\vspace{-5pt}
%raybuck2021hemem

%\textbf{Workloads} \textcolor{shangye}{In our platform, five different representative workloads are used for evaluation, including graph-related benchmark (BFS,SSSP, PageRank), a workload related to Monte Carlo neutron transport algorithm computation (XSBench), a retrieval related to in-memory index (Btree). These workloads are commonly used in the field of managing huge pages and evaluating tired memory systems \cite{10.1145/3582016.3582063,10.1145/3600006.3613167,guo2017smartmd,kumar2021radiant,kwon2016coordinated, raybuck2021hemem,panwar2019hawkeye,ram2021trident,achermann2020mitosis,alverti2020enhancing,panwar2021fast}. In Table \ref{tab:benchmark}, we can see various workloads and their descriptions. }

\vspace{-5pt}

\subsection{Model Accuracy} 
\vspace{-5pt}
We change the fast memory size, and evaluate the model accuracy. We use TPP for page migration during the evaluation. For each benchmark, we run it with the whole RSS in fast memory, measure its performance (denoted by $x$), and profile the workload to build a configuration vector. Then we reduce the fast memory size to various values, re-run the benchmark, and measure its performance. For each fast memory size, we measure the performance of the benchmark (denoted by $y$), and calculate the performance difference ($pd$) using $x$ as the baseline, i.e., $pd = (y-x)/x$. We use the configuration vector to query the performance database and find the performance results collected using various fast memory sizes. For a given configuration vector and a fast memory size, from the performance database we get the performance (denoted by $x'$) using fast memory only and the performance (denoted by $y'$) using the given fast memory size. Then we calculate the performance difference ($pd'$) using $x'$ as the baseline, i.e., $pd' = (y'-x')/x'$. We report $|pd' - pd|/pd$ in Table~\ref{tab:performanceacc}.
\vspace{-3pt}
In general, the table shows high model accuracy: the model prediction error is consistently less than 10\%. This indicates that the performance database and micro-benchmark reliably capture the performance difference (in terms of percentage) between using the fast memory only and using smaller fast memory. Furthermore, we notice a trend that the model prediction error tends to increase when the fast memory size is smaller (e.g., 85\% of the full fast-memory size). However, saving 15\% (or even higher) of fast memory size usually leads to large performance loss, and we do not see the cases that tolerates such a large loss.

\begin{figure}[ht]
    \begin{minipage}{0.45\columnwidth} 
        \centering
        \includegraphics[width=\linewidth]{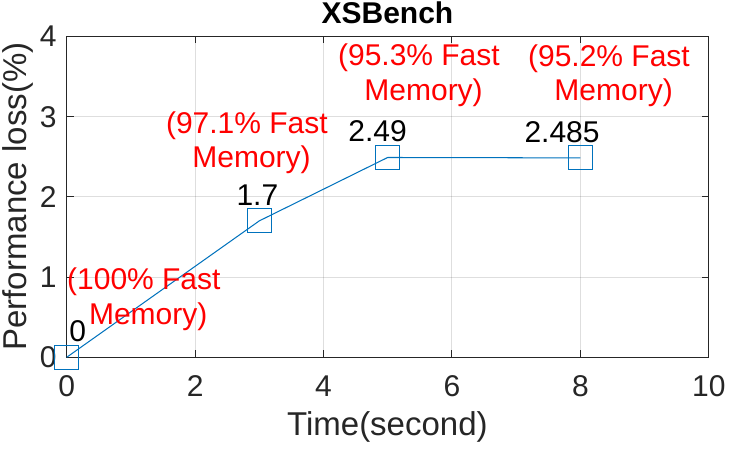} 
        \vspace{-5pt}
        \caption{Fast memory variance: XSBench}
        \label{fastmemoryvarianceXSBench}
    \end{minipage}
    \hfill
    \begin{minipage}{0.45\columnwidth}
        \centering
               \includegraphics[width=\linewidth]{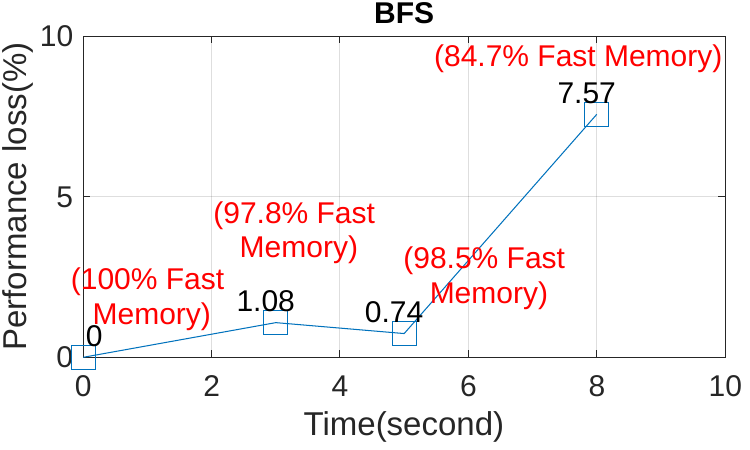} 
               \vspace{-5pt}
        \caption{Fast memory variance: BFS}
        \label{fastmemoryvarianceBFS}
    \end{minipage}
\end{figure}

\vspace{-5pt}

\begin{figure}[ht]
    \begin{minipage}{0.45\columnwidth} 
        \centering
        \includegraphics[width=\linewidth]{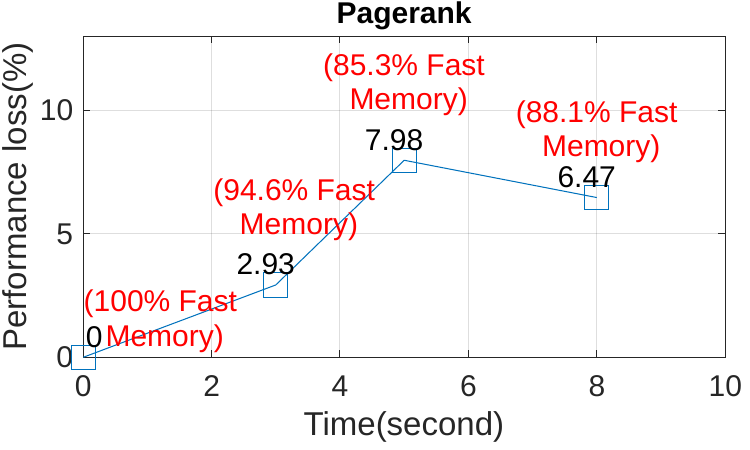}
        \vspace{-5pt}
        \caption{Fast memory variance: PageRank}
        \label{fastmemoryvariancepagerank}
    \end{minipage}\hfill
    \begin{minipage}{0.45\columnwidth}
        \centering
        \includegraphics[width=\linewidth]{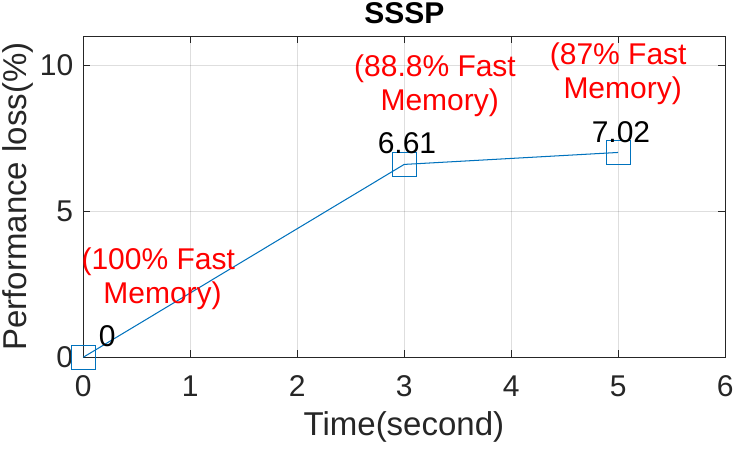}
        \vspace{-5pt}
        \caption{Fast memory variance: SSSP}
        \label{fastmemoryvariancesssp}
    \end{minipage}
\end{figure}
\vspace{-5pt}
\begin{comment}
\begin{figure}
	\begin{minipage}[t]{0.5\linewidth}
		\centering
		\includegraphics[width=2.2in]{figures/evaluation2pagerank.pdf}
		\caption{FM variance: Pagerank}
		\label{fastmemoryvariancepagerank}
	\end{minipage}
	\begin{minipage}[t]{0.5\linewidth}
		\centering
		\includegraphics[width=2.2in]{figures/evaulation2sssp.pdf}
		\caption{FM variance: SSSP}
		\label{fastmemoryvariancesssp}
	\end{minipage}
\end{figure}
\end{comment}

\begin{comment}
\begin{figure}    
\centering     
\includegraphics[width=0.45\textwidth]{figures/evaluation2pagerank.pdf}
\caption{Fast memory variance and performance accuracy}
\label{fig5:fastmemoryvariance}
\end{figure}
\begin{figure}    
\centering     
\includegraphics[width=0.5\textwidth]{figures/evaulation2sssp.pdf}
\caption{Fast memory variance and performance accuracy}
\label{fig6:fastmemoryvariance}
\end{figure}
\end{comment}

\begin{figure}[ht]
    \begin{minipage}{0.47\columnwidth} 
        \centering
        \includegraphics[width=\linewidth]{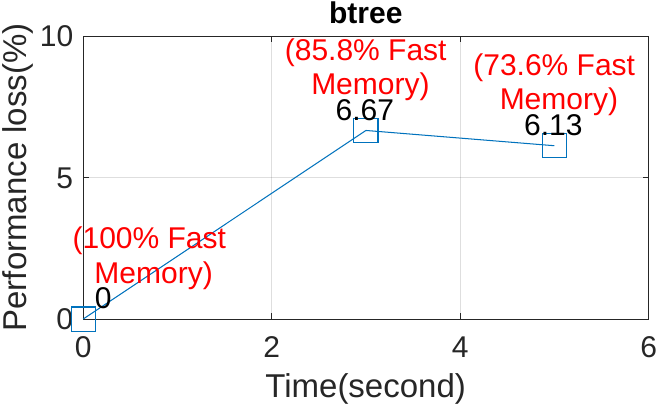}
        \vspace{-10pt}
        \caption{Fast memory variance: Btree}
        \label{fastmemoryvariancebtree}
    \end{minipage}
    \hfill
    \begin{minipage}{0.47\columnwidth}
        \centering
        \includegraphics[width=\linewidth]{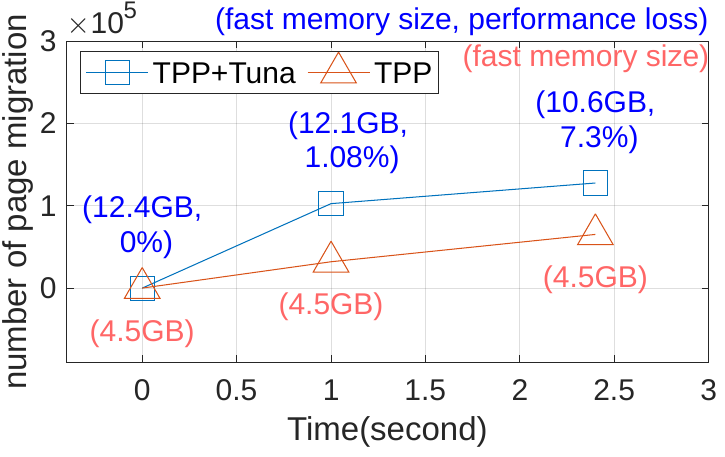}
        \vspace{-10pt}
        \caption{Performance of TPP with and without \name for bfs.}
        \label{comparisontpp}
    \end{minipage}
\end{figure}

\vspace{-13pt}

\begin{table}[!t]
    \begin{center}
        \begin{threeparttable}
            \small
            \setlength{\tabcolsep}{4pt} % Adjust column spacing
            
            \caption{Model prediction error. ``FM'' is the percentage of full fast-memory size.  ``MA'' shows the model prediction error.}
            \vspace{-10pt}
            \begin{tabular}{lccccccccccc}
                \multicolumn{1}{c}{} & \textit{\textbf{FM}} & \textbf{99\%} & \textbf{98\%} & \textbf{97\%} & \textbf{96\%} & \textbf{95\%} & \textbf{88\%} & \textbf{85\%} \\ \hline
                \small{\textbf{SSSP} }     & \textit{\textbf{\small{MA}}} & 0.6\% & 1.2\% & 1.7\% & 2.3\% & 2.9\% & 6.3\% & 8.0\% \\ \hline
                \textbf{BFS}       & \textit{\textbf{MA}} & 0.5\% & 1.0\% & 1.6\% & 2.1\% & 2.6\% & 5.7\% & 7.3\% \\ \hline
                \textbf{\small{XSBench}}   & \textit{\textbf{MA}} & 0.4\% & 1.2\% & 1.8\% & 2.3\% & 2.9\% & 6.3\% & 8.1\% \\ \hline
                \textbf{\small{PageRank}} & \textit{\textbf{MA}} & 0.2\% & 0.5\% & 0.8\% & 1.1\% & 2.9\% & 6.4\% & 7.9\% \\ \hline
                \textbf{Btree}     & \textit{\textbf{MA}} & 0.4\% & 0.9\% & 1.4\%  & 1.9\% & 2.5\% & 5.8\%  & 6.8\%
            \end{tabular}
            \label{tab:performanceacc}
            % ... Tablenotes if needed
        \end{threeparttable}
    \end{center}
\end{table}
\vspace{-15pt}

\subsection{Tuning Fast Memory Size}
\vspace{-5pt}
We apply \name to all benchmarks. We use 5\% as the performance loss target. Figures~\ref{fastmemoryvarianceXSBench}-\ref{fastmemoryvariancebtree} show the memory saving and performance loss at runtime. The performance loss is the application performance loss after changing the fast memory size but before changing it again. We change the fast memory size every 2.5 seconds. \textcolor{check}{The overall performance loss (not shown in the figures) for XSBench, BFS, PageRank, SSSP, and Btree is 1.8\%, 2\%, 4.6\%, 4.7\%, and 4.6\%,  respectively.} The performance loss is consistently within the performance loss target. 

\name effectively leads to fast memory saving (up to 16\% shown in Figure~\ref{fastmemoryvariancebtree} for Btree). At runtime, the performance loss can violate the performance target for a time duration, but since the performance loss is within the target most of the time, the overall performance is still within the target. 
\vspace{-1pt}
%%%%\textcolor{shangye}{For evaluation and Verify the relationship between performance loss and fast memory size, we use the dynamic tuning fast memory size to validate the results, in Figure\ref{fastmemoryvarianceXSBench}, \ref{fastmemoryvarianceBFS},\ref{fastmemoryvariancepagerank},\ref{fastmemoryvariancesssp} and \ref{fastmemoryvariancebtree}, given a user-specified performance loss target, 5\%, each workload will start running with fast memory only condition, then in every two seconds as a time interval, we will tune the fast memory size by decreasing based on the demand, then the workload will run with the tuned fast memory size. With a few times adjustments for fast memory size, the desired value of fast memory can be achieved meanwhile meeting the specified performance loss. Finally, We can find the fast memory variance and the performance for the five different workloads with runtime.} 

We look into the performance difference between TPP and TPP+\name. 
Figure~\ref{comparisontpp} shows the number of page migrations and fast memory saving at runtime. TPP is not designed to save fast memory. With \name, as the fast memory size is changed, we see the changes in the number of page migrations managed by TPP.  Such changes allows the application to save fast memory size without large performance loss. 
\vspace{-2pt}
%{TPP only uses the demotion function to demote cold pages to slow memory without actively changing the fast memory size, thus we use our design and model to trigger TPP for changing the fast memory size to improve the number of page migrations and meet user requirements during runtime. To validate the effectiveness of the model, we use workload BFS to compare TPP with our design in two conditions(TPP and TPP plus our tuning fast memory size) to evaluate the contribution of the two different methods to the number of page migrations. Figure \ref{comparisontpp} shows the results of the comparison with TPP and with our design during runtime, the workload BFS starts running with fast memory only, and then we change the fast memory size with the specific time interval. It clearly shows that the number of page migrations with TPP does not change too much, while the number of page migrations with our method shows a good contribution to the number of page migrations, which is much higher than TPP. The evaluation proves that our design can actively promote page migration. }
\vspace{-5pt}
\subsection{Sensitivity Study}
\vspace{-5pt}
\begin{table}
    \centering
    \caption{Performance sensitivity to performance loss target}
    \vspace{-10pt}
    \begin{tabular}{lccc}
        \toprule
        \textbf{Performance loss target} & \textbf{5\%} & \textbf{10\%} & \textbf{15\%} \\
        \midrule
        \textbf{SSSP Fast Memory Saving}   & 9\% & 18\% & 27\% \\
        \textbf{SSSP Performance Loss}   & 4.6\%  & 9.6\%  & 15.1\% \\
        \bottomrule
    \end{tabular}
      \label{tab:sensitivity}
      \vspace{-10pt}
\end{table}

%\textcolor{shangye}{To evaluate the sensitivity of performance loss target to our design, In Table \ref{tab:sensitivity}, we evaluated the workload sssp with three different performance loss targets which are 5\%, 10\%, and 15\%. With each of the performance loss targets, we can find the actual performance loss workload SSSP is always near or smaller than the user-specific performance loss target. The results prove that the system is not sensitive to the performance loss target, which means the performance loss target can easily control the system. }

% \vspace{-5pt}
\textbf{Sensitivity to performance loss target.} We use SSSP and three different performance loss targets shown in Table~\ref{tab:sensitivity} for evaluation. \name allows us to control the performance loss within the target except the case of 15\% as the target. In this case, the model prediction error can be relatively high (shown in Table~\ref{tab:performanceacc}), which leads to the slight violation. 

\textbf{Sensitivity to the frequency of tuning fast memory size.} By default, \name tunes the fast memory size every 2.5s. We change this frequency to 0.5s, 1s, and 5s. We see obvious tradeoff between memory saving and performance loss. Take SSSP as an example. Using a smaller frequency (0.5s), \name saves memory by up to 25\%, but suffers from 17\% performance loss. Using a larger frequency (5s), \name saves fast memory by up to 2\%, but causes ignorable performance loss (3\%). We choose 2.5s to strike a balance between memory saving and performance loss.  
\vspace{-5pt}
%%{To evaluate the sensitivity of performance loss target to our design, In Table \ref{tab:sensitivity}, we evaluated the workload sssp with three different performance loss targets which are 5\%, 10\%, and 15\%. With each of the performance loss targets, we can find the actual performance loss workload SSSP is always near or smaller than the user-specific performance loss target. The results prove that the system is not sensitive to the performance loss target, which means the performance loss target can easily control the system.   }
\vspace{-5pt}
\section{Related Work}
\vspace{-10pt}
\textbf{Page management for tiered memory.} 
Many efforts focus on page profiling and migration on tiered memory ~\cite{atc21:autotiering, 10.1145/3582016.3582063, 10.1145/3477132.3483550, 10.1145/3600006.3613167, Agarwal:2017:TAP:3037697.3037706, asplos19:yan, kumar2021radiant, liu2019hierarchical, heo2020adaptive, sha2023vtmm, li2019thinking, maruf2022multi, kannan2020efficient, choe2020study, zheng2019ziggurat, ren2023hm, lee2022optimizing, wang2019supporting, ni2023tmc, 10.1145/3582016.3582031}. TPP~\cite{10.1145/3582016.3582063} is a recent solution from industry. TPP decouples the memory allocation and reclamation logic to avoid promotion failures, and introduces a lightweight page reclamation mechanism. MEMTIS~\cite{10.1145/3600006.3613167} uses histogram to characterize access distribution of pages to identify 
the hot data set for promotion. MEMTIS also dynamically splits huge pages to avoid co-location of hot and cold pages to improve the utilization of fast memory. HeMem~\cite{10.1145/3477132.3483550} uses performances counters to sample memory accesses across pages. Thermostat~\cite{Agarwal:2017:TAP:3037697.3037706} estimates the performance degradation due to placing a particular page in slow memory, based on which it introduces a huge-page-aware hot/cold page classification for tiered memory.  However, none of the existing efforts tune the fast memory size.

\textbf{System configuration tuning.} Many efforts tune system configurations for high performance~\cite{wei2014self,weikum2002self,dias2005automatic,zhang2022towards, zhu2017bestconfig, zhang2019end, li2019qtune, kunjir2020black,fekry2020tune,fekry2020tuneful,makiyama2015text,jain2018database,zolaktaf2020facilitating, zhang2021restune,van2017automatic,duan2009tuning}. 
The tuning methods can be generally classified into two types~\cite{zhang2022towards}: (machine) learning-based and rule-based. ResTune~\cite{zhang2021restune} is a learning-based system that applies the Bayesian Optimization (BO) method to tune configuration knobs of database management systems (DBMS). Qtune~\cite{li2019qtune} and CDBTune~\cite{zhang2019end} use the reinforcement learning (RL) to tune configuration knobs of DBMS.  ReIM \cite{kunjir2020black} tunes the memory allocation for applications running on distributed data processing systems. ReIM builds an analytical solution that captures the interactions between resource manager (e.g., YARN), containers, and applications. Tuneful \cite{fekry2020tune,fekry2020tuneful}  tunes the configurations of data analytics frameworks (e.g., Spark). They use incremental sensitivity Analysis and the Bayesian optimization to identify near-optimal configurations. 

The rule-based method uses predefined rules or heuristics to optimize the configuration settings. For example, BestConfig~\cite{zhu2017bestconfig} tunes the configurations for data analytic systems and databases within a resource limit, using two heuristics (including a divide-and-diverge sampling method and the recursive bound-and-search algorithm). The rule-based method, although providing a straightforward and interpretable approach for system configuration tuning, heavily relies on the assumptions in their heuristics.
\vspace{-5pt}

\vspace{-5pt}
\section{Conclusions}
\label{sec:conclusion}
\vspace{-10pt}
Deciding fast memory size is critical to maximize the benefits of tiered memory for memory-consuming applications. However, optimizing the fast memory size is less studied as an optimization target. In this paper, we demonstrate the potential of saving fast memory in combination of page migration. We also introduce general modeling techniques to determine the fast memory size. We introduce a system  \name that automatically decides fast memory size at runtime based on offline micro-benchmarking-based modeling. 
%\vspace{-10pt}

%\bibliography{\jobname}
\bibliographystyle{plain}
\bibliography{li,shangye}

%%%%%%%%%%%%%%%%%%%%%%%%%%%%%%%%%%%%%%%%%%%%%%%%%%%%%%%%%%%%%%%%%%%%%%%%%%%%%%%%
\end{document}